\documentclass[12pt]{article}

\usepackage{latexsym}

\begin{document}

\title{Self-similar collapse of a scalar field in dilaton gravity and 
critical behaviour}

\author{
     Stoytcho S. Yazadjiev \thanks{E-mail: yazad@phys.uni-sofia.bg}\\
{\footnotesize  Department of Theoretical Physics,
                Faculty of Physics, Sofia University,}\\
{\footnotesize  5 James Bourchier Boulevard, Sofia~1164, Bulgaria }\\
}

\date{}

\maketitle

\begin{abstract}
We present new analytical self-similar solutions describing a
collapse of a massless scalar field in scalar-tensor theories. The
solutions exhibit a type of critical behavior. The black hole
mass for the near critical evolution  is analytically  obtained
for several scalar-tensor theories and the critical exponent is
calculated. Within the framework of the analytical models we
consider it is found that the black hole mass law for some
scalar-tensor theories is of the form
$M_{BH}=f(p-p_{cr})(p-p_{cr})^\gamma$ which  is slightly
different from the general relativistic law $M_{BH}=const
(p-p_{cr})^\gamma$.
\end{abstract}

\section{Introduction}

Critical phenomena in gravitational collapse have attracted much
interest since the pioneering work by Choptuik \cite{C1} and they
are considered to be one of the greatest successes of the
numerical relativity. The essence of the critical phenomena
consists in the fact that, just at the threshold of the black
hole formation, the field dynamics becomes simple and exhibits
discrete or continuous self-similarity despite the nature and
non-linearity of the collapsing matter. The critical solution
separates the solutions with a black hole formed from those
without a formation of a black hole. The mass of a black hole
formed in near critical collapse obeys the power law

\begin{equation}
M_{BH}=C(p-p_{cr})^\gamma
\end{equation}

where the parameter $p$ describes the strength of the initial data
and $p_{cr}$ is its  critical value.The exponent $\gamma$ is, in
general, matter dependent and $C$ is a constant which depends on
the initial field configuration. For detail discussion of the
critical phenomena we refer the reader to the Gundlach's review
\cite{G} and references therein.

Parallel to the numerical investigations , an analytical approach
has been undertaken  in order to understand more deeply the
critical collapse. Although fascinating, this task was proved to
be very difficult. The analytical approach is based on approximate
(perturbative) techniques   or exact toy models which cannot be
considered as completely realistic and satisfactory.
Nevertheless, the analytical methods, and especially the exact
solutions provide valuable independent insight into the subject.
It was realized by Brady \cite{B} and Oshiro, Nakamura and
Tomimatsu \cite{ONT} that the one parameter family of exact
self-similar real massless scalar-field solutions first
discovered by Roberts\footnote{It turns out that the original
Roberts's solution as presented in \cite{R} solves the field
equations only for one nontrivial case and represents just a
class of measure zero among the solutions to the spherically
symmetric homothetic Einstein-scalar field equations given in
\cite{B}, \cite{ONT} and \cite{SUS}. For detail discussion see
\cite{BURKO} where the Roberts's solution is generalized. I would
like to thank L. Burko for pointing me this out. } \cite{R} can be
applied to explain some basic features of the critical collapse.
Solutions from this family with $p>1$ describe the formation of a
black hole and to the leading order around the critical value
$p_{cr}=1$ the black hole mass turns out to be
$M\sim(p-1)^{1/2}$. The critical exponent in this toy model is
therefore $\gamma=1/2$ and is slightly different from the
numerically calculated value $\gamma=0.37$. In order to gain as
much as possible insight into the critical phenomena, some exact
toy models in three dimensions were also considered \cite{GAR},
\cite{CF}.

Critical collapse within the scalar-tensor theories  has
attracted some interest, too \cite{C2},\cite{CS}. There were also
some analytical works on the critical collapse in Brans-Dicke
theory and  the theory with conformal coupling
\cite{CS},\cite{DOCT},\cite{DO}. Using conformally transformed
Roberts's solutions the authors were able to find analytically
the black hole mass power law and to extract the critical
exponent which turns out to be $ \gamma =1/2 \pm 1/2\sqrt{2\omega
+ 3}$  and  $\gamma = 0.21$ for the Brans-Dicke and the theory
with conformal coupling, respectively.

However, there are serious  omissions in the works treating
analytically critical collapse in scalar-tensor theories. First,
only two theories were considered. Second, the authors have
considered simplified models where the matter fields are absent
(dilaton-vacuum solutions) i.e. the collapsing matter has been
taken to be the dilaton field. The aim of the present work is to
fill in this gap studying analytically the critical collapse  of
a massless real scalar field  within the scalar-tensor theories
of  gravity. Our study is based on exact solutions which
generalize Roberts's solution in the presence of a dilaton field.

\section{Exact self-similar solutions }

The general form of the extended gravitational action in
scalar-tensor theories is

\begin{eqnarray} \label{JFA}
S = {1\over 16\pi G_{*}} \int d^4x \sqrt{-{\tilde
g}}\left({F(\Phi)\tilde R} - Z(\Phi){\tilde
g}^{\mu\nu}\partial_{\mu}\Phi
\partial_{\nu}\Phi  \right. \nonumber  \\ \left. -2 U(\Phi) \right) +
S_{m}\left[\Psi_{m};{\tilde g}_{\mu\nu}\right] .
\end{eqnarray}

Here, $G_{*}$ is the bare gravitational constant, ${\tilde R}$ is
the Ricci scalar curvature with respect to the space-time metric
${\tilde g}_{\mu\nu}$. The dynamics of the scalar field $\Phi$
depends on the functions $F(\Phi)$, $Z(\Phi)$ and $U(\Phi)$. In
order for the gravitons  to carry positive energy the function
$F(\Phi)$ must be positive. The nonnegativity of the energy of
the dilaton field requires that $2F(\Phi)Z(\Phi) +
3[dF(\Phi)/d\Phi]^2 \ge 0$. The action of matter depends on the
material fields $\Psi_{m}$ and the space-time metric ${\tilde
g}_{\mu\nu}$.

However, it is much more convenient from a mathematical point of
view to analyze the scalar-tensor theories with respect to the
conformally  related Einstein frame  given by the metric:

\begin{equation}\label {CONF1}
g_{\mu\nu} = F(\Phi){\tilde g}_{\mu\nu} .
\end{equation}

Further, let us introduce the scalar field $\varphi$ (the so
called dilaton) via the equation

\begin{equation}\label {CONF2}
\left(d\varphi \over d\Phi \right)^2 = {3\over
4}\left({d\ln(F(\Phi))\over d\Phi } \right)^2 + {Z(\Phi)\over 2
F(\Phi)}
\end{equation}

 and define

\begin{equation}\label {CONF3}
{\cal A}(\varphi) = F^{-1/2}(\Phi) \,\,\, ,\nonumber \\
2V(\varphi) = U(\Phi)F^{-2}(\Phi) .
\end{equation}

In the Einstein frame action (\ref{JFA}) takes the form

\begin{eqnarray}
S= {1\over 16\pi G_{*}}\int d^4x \sqrt{-g} \left(R -
2g^{\mu\nu}\partial_{\mu}\varphi \partial_{\nu}\varphi -
4V(\varphi)\right) \nonumber \\ + S_{m}[\Psi_{m}; {\cal
A}^{2}(\varphi)g_{\mu\nu}]
\end{eqnarray}

where $R$ is the Ricci scalar curvature with respect to the
Einstein metric $g_{\mu\nu}$.

In what follows we shall consider the case $U(\Phi)=0$ and the
matter source  will be a massless real scalar field $\sigma$.

The starting point in our considerations is the Roberts's
solution which is written as

\begin{eqnarray}
ds^2 &=& -dudv + r^2(u,v)d\Omega^2 ,\\
\sigma_{0} &=& \pm {1\over 2}\ln\left({u - (1+p)v \over u -
(1-p)v }\right),
\end{eqnarray}

where $$r^2(u,v) ={1\over 4}[u - (1-p)v][u - (1+p)v]$$ and  $p>0$
is an arbitrary constant.

Dilaton-vacuum solutions can be easily obtained via the conformal
transformation

\begin{eqnarray} \label{DVS}
F^{-1}(\Phi_{dv}(u,v)) = {\cal A}^2(\varphi)|_{\varphi=\sigma_{0}(u,v) }, 
\\
d{\tilde s}^2_{dv}= F^{-1}(\Phi_{dv}(u,v)) ds^2.
\end{eqnarray}

This simple procedure, however, is not applicable when a matter
source is present. In order to construct exact scalar-tensor
solutions describing a collapsing scalar field we shall employ
the solution generating methods developed in \cite{Y}. Roberts's
solution is taken as a seed solution.

The general form of the scalar-tensor solutions is given by

\begin{eqnarray}
F^{-1}[\Phi(u,v)] &=& F^{-1}[\Phi(\sigma_{0}(u, v))] = {\cal 
A}^2[\varphi(\sigma_{0}(u,v))], \\
d{\tilde s}^2 &=& F^{-1}[\Phi(\sigma_{0})]
ds^2 = -F^{-1}[\Phi(\sigma_{0})]dudv + {\tilde r}^2(u,v)d\Omega^2 ,\\
\sigma(u,v) &=& \sigma(\sigma_{0}(u,v)),
\end{eqnarray}

where

\begin{equation}
{\tilde r}^2(u,v) = F^{-1}[\Phi(\sigma_{0})] r^2(u,v)
\end{equation}

and the explicit form of the functions $F^{-1}(\Phi(\sigma_{0}))$
and $\sigma(\sigma_{0})$ depends on the particular scalar-tensor
theory. Below we consider several examples of scalar-tensor
theories which qualitatively cover the general case and have
solutions representable  in a closed analytical form.

The solutions above are continuously self-similar since they admit
the homothetic Killing vector
\begin{equation}
\xi = u{\partial\over \partial u} + v{\partial\over \partial v}.
\end{equation}

The Ricci scalar curvature ${\tilde R}$ is given by

\begin{equation}
{\tilde R}= F(\Phi(\sigma_{0}))\left(1 + {3\over 2} {d^2\over
d\sigma^2_{0}} \ln\left[F(\Phi(\sigma_{0}))\right]  -{3\over 4}
\left({d\over d\sigma_{0}} \ln\left[F(\Phi(\sigma_{0}))\right]
\right)^2\right) R
\end{equation}

where

\begin{equation}
R= {p^2 uv \over 2r^4(u,v)}
\end{equation}

is the Ricci scalar curvature for the Roberts's solution.

For all solutions we present here the conformal factor
$F^{-1}(\Phi)$ satisfies $F^{-1}(\Phi) \ge \lambda^2 >0$ where
$\lambda$ is a constant.  Therefore, the conformal factor is
everywhere regular except at $r(u,v)=0$ where the only
singularity\footnote{For all scalar-tensor solutions presented
here it can be checked that ${\tilde R}$ could become singular
only for $R$. } of the Ricci scalar curvature is located. Thus we
can conclude that the causal structure of the scalar-tensor
solutions is the same as the Roberts's solution. The singularity
$r(u,v)=0$ is given by $u=(1-p)v$ and $u=(1+ p)v$ for $v>0$ and
$v<0$, respectively. For $0<p<1$ the singularity is timelike
while it is null for $p=1$. When $p>1$ the singularity is
spacelike in the region $v>0$ and timelike for $v<0$.

In all cases we consider, an apparent horizon exists only for
$p>1$ and it surrounds the spacelike singularity while the
timelike singularity is naked. In order to get rid of the naked
singularity we shall proceed as in the case of the Roberts's
solution. We can replace the region $v<0$ by a flat spacetime
smoothly matched to the region $v>0$ across the $v=0$ line where
the energy flux vanishes in both the Jardan and Einstein frame.

The fact that the spacelike singularity is hidden behind and
apparent horizon signals the formation of a black hole. That is
why we shall consider bellow the case $p>1$ in the limit $p\to 1$.

With regard to the case $0<p<1$, we can, just as for the Roberts's
solutoin, remove the timelike singularity by replacing the region
$u>0 $ and  $v>0$ by a flat spacetime. For $0<p<1$ the initial
conditions are not strong enough to form a black hole.

The above considerations show that the parameter $p$ is a control
parameter and its critical value is $p_{cr}=1$. Thus the
presented solutions can be divided into three classes:
subcritical ($0<p<1$) ,  critical ($p=1$) and supercritical
($p>1$).

\subsection {Brans-Dicke theory}

Brans-Dicke theory is described  by the functions $F(\Phi)=\Phi$
and $Z(\Phi)= \omega/\Phi$ where $\omega$ is a constant. Here we
shall consider the cases with $\omega >0$.

Solution generating transformations \cite{Y}  give the following
scalar-tensor solution:

\begin{eqnarray}
\Phi^{-1}(u,v) &=&  \lambda^2 \cosh^2\left({\sigma_{0}(u,v)
 \over \sqrt{3 + 2\omega}}\right), \\
\sigma(u,v) &=& {\sqrt{3 + 2\omega} \over \lambda}
\tanh\left({\sigma_{0}(u,v)
 \over \sqrt{3 + 2\omega}} \right),
\end{eqnarray}

where $0<\lambda^2 <\infty$.

The apparent horizon exist only for $p>1$ and is located at

\begin{equation}
u \cong     {2(1-p) \over 1 \pm {1\over \sqrt{3 + 2\omega}}} v .
\end{equation}

Here and below, in order to present the results in compact and
clear form the location of the apparent horizon is determined for
$(p-1)<<1$. The effective gravitational mass inside the apparent
horizon is

\begin{equation}
{\tilde M}_{AH} = {1\over 2} {\tilde r}_{AH} \cong {\sqrt{2}\over
8} |\lambda| \left({1 \mp {1\over \sqrt{3 + 2\omega}} \over 1 \pm
{1\over \sqrt{3 + 2\omega}} } \right)^{1/2} \left(p-1\right)^{1/2
- 1/2\sqrt{3 + 2\omega}}  v.
\end{equation}

As well know the self-similar spacetimes are not asymptotically
flat and the black hole mass diverges when $v\to \infty$. The
standard way to get rid of this problem is to truncate the
spacetime at some $v_{*}>0$ and to add an asymptotically flat
region. We shall accept from now on that this has been done.

What is important is that the black hole mass obeys the power law.
Hence the analytically calculated critical exponent is found to
be $\gamma= {1\over 2} - {1\over 2\sqrt{3 + 2\omega}}$. In
contrast to the dilaton-vacuum case  where $\gamma_{dv}= 1 \pm
{1\over 2\sqrt{3 + 2\omega}}$ \cite{CS}, \cite{DO} , in the
presence of  a massless scalar field as a matter source, only the
minus sign is allowed.

\subsection{Theory with conformal coupling}

This theory is described by the functions $F(\Phi)= 1 -{1\over
6}\Phi^2$ and $Z(\Phi)=1$. The explicit functions in the case
under consideration are:

\begin{eqnarray}
F^{-1}(\Phi(u,v)) = 1 +
(1-\lambda^2) \sinh^2\left({\sigma_{0}(u,v)\over \sqrt{3}}\right) , \\
\sigma(u,v) = \sqrt{3} \tanh^{-1}\left(\lambda
\tanh\left({\sigma_{0}(u,v)\over \sqrt{3}}\right) \right),
\end{eqnarray}

where the parameter $\lambda$ is restricted by $0<\lambda^2 <1$.

The apparent horizon exists for $p>1$ and is located at

\begin{equation}
u \cong {2(1-p) \over  1 \pm  {1\over \sqrt{3}} } v
\end{equation}

The mass inside the apparent horizon is

\begin{equation}
{\tilde M}_{AH} \cong {\sqrt{2}\over 4} \left(1-\lambda^2
\right)^{1/2} \left({1 \mp {1\over \sqrt{3}} \over 1 \pm {1\over
\sqrt{3}} } \right)^{1/2} \left(p -1\right)^{1/2 - 1/2\sqrt{3}}
v_{*}.
\end{equation}

The critical exponent is $\gamma = {1\over 2} - {1\over
2\sqrt{3}}$ and coincides with the previously calculated critical
exponent $\gamma_{dv} \approx 0.21$ for the dilaton-vacuum
solutions \cite{DOCT}.

\subsection{$F(\Phi)=\Phi$ and $Z(\Phi)= (\Phi^2 -3\Phi +3)/2\Phi(\Phi-1)$ 
theory }

The explicit form the functions $\Phi(\sigma_{0})$ and
$\sigma(\sigma_{0})$ is given by

\begin{eqnarray}
\Phi^{-1}(u,v) &=&
{\lambda^2 \over \lambda^2 + (1-\lambda^2)\sin^2\left(\lambda 
\sigma_{0}(u,v)\right)} ,\\
\sigma(u,v) &=& {1 + \lambda^2 \over 2\lambda }\sigma_{0}(u,v)  -
{1-\lambda^2\over 4\lambda^2}\sin\left(2\lambda \sigma_{0}(u,v)
\right).
\end{eqnarray}

The range of the parameter $\lambda$ is $0<\lambda^2 <1$.

The apparent horizon is located at

\begin{equation}
u \cong  { 2(1- p) \over  1 \pm D_{\lambda}(p)} v
\end{equation}

where

\begin{equation}
D_{\lambda}(p) = {1\over 2} { \lambda
(1-\lambda^2)\sin\left(\lambda \ln(p-1) \right) \over \lambda^2 +
(1-\lambda^2)\sin^2\left({1\over 2}\lambda \ln(p-1)\right) }
\end{equation}

and $D_{\lambda}^2(p)<1$.

The mass inside the apparent horizon is given by

\begin{equation}
{\tilde M}_{AH} \cong {2^{1/2} \over 4} {\mid\lambda \mid \over
\sqrt{ \lambda^2 + (1-\lambda^2)\sin^2\left({1\over 2}\lambda
\ln(p-1)\right)} } \left({1 \mp D_{\lambda}(p)  \over 1 \pm
D_{\lambda}(p)}\right)^{1/2} \,(p-1)^{1/2} v_{*}  .
\end{equation}

We see something interesting. The power law $M=const
(p-p_{cr})^\gamma$ observed within the general relativity is
slightly modified: $M =f(p-p_{cr})(p-p_{cr})^\gamma$. It is also
seen that the black hole mass exhibits " damping oscillations" in
the control parameter $p$ when $p\to 1$.

If we define the critical exponent as $\gamma = \lim_{p\to 1}
\ln(M)/\ln(p-1)$ we obtain $\gamma = 1/2$ for the case under
consideration.

It is also interesting to find the black hole mass law for the
dilaton-vacuum collapse. Using solution generating formulas
(\ref{DVS}) one can show that

\begin{eqnarray}
{\tilde M}^{dv}_{AH} \cong {1 \over \sqrt{2} } {(p-1)^{1/2} \over
\mid\ln(p-1)\mid } v_{*} .
\end{eqnarray}

Hence we find $\gamma_{dv} = \lim_{p\to 1} \ln(M)/\ln(p-1) = 1/2$.

We have checked that the black hole mass for the near critical
collapse in the Barker's theory ($F(\Phi)=\Phi$ and $Z(\Phi)=
(4-3\Phi)/2\Phi(\Phi-1)$) exhibits "damping oscillations" in the
critical parameter, too. Since the Barker's case is similar to
the case under consideration we do not present detail
calculations.

\subsection{$F(\Phi)=\Phi$ and $  Z(\Phi)= (1-3a^2 \Phi)/2a^2 \Phi^2 $ 
theory}

Here the parameter $a$ satisfies $a>0$.

In the case under consideration we have

\begin{eqnarray}
\Phi^{-1}(u,v) = \left(1 + a \sigma_{0}(u,v)\right)^2 + \lambda^2  , \\
\sigma(u,v) = {1\over a}\arcsin\left({\lambda \over \sqrt{(1 + a
\sigma_{0}(u,v))^2 + \lambda^2  }} \right) .
\end{eqnarray}

Here, $\lambda$ runs $0< \lambda^2 < \infty$.

The location of the apparent horizon is given by

\begin{equation}
u \cong 2(1-p)v .
\end{equation}

The mass inside the apparent horizon is

\begin{equation}
{\tilde M}_{AH} \cong {\sqrt{2} \over 4}a \mid \ln(p-1)\mid
(p-1)^{1/2} v_{*} .
\end{equation}

Hence, for the critical exponent we have $\gamma = \lim_{p\to 1}
\ln(M)/\ln(p-1) = 1/2$.

Respectively, for the dilaton-vacuum collapse we obtain

\begin{equation}
{\tilde M}^{dv}_{AH} \cong {\sqrt{2} \over 4}a \mid \ln(p-1)\mid
(p-1)^{1/2} v_{*}
\end{equation}

and therefore $\gamma_{dv}=\gamma=1/2$.

\section{Conclusion}

In the present work we have generalized the general relativistic
Roberts-Brady-Oshiro- Nakamura-Tomimatsu (RBONT) model in the case
of scalar-tensor theories of gravity. We have presented new exact
self-similar solutions describing a collapse of a massless
scalar  field in scalar-tensor theories. The solutions exhibit a
type of critical behavior discussed by Choptuik. Three possible
evolutions are distinguished: subcritical, critical and
supercritical. For supercritical evolution the black hole mass
low has been found and the critical exponent has been extract for
several scalar- tensor theories which cover qualitatively the
general case. It is interesting to note that for some
scalar-tensor theories within the analytical models we consider,
the black hole mass law in the near critical collapse is of the
form $M=f(p-p_{cr})(p-p_{cr})^\gamma$ where $f(p-p_{cr})$ is a
function depending on the particular scalar-tensor theory. One
sees that this law is slightly different compared to the general
relativistic version $M=const (p-p_{cr})^\gamma$. It is also
interesting to note that, in some cases, the black hole mass
exhibits "damping oscillations" in the limit $p \to p_{cr}$.

As the RBONT model our models can not be considered as completely
realistic but they are satisfactory enough in explaining
analytically some basic features of the critical collapse.

\bigskip

\section*{Acknowledgments}

This work was supported in part by Sofia University Grant No
3429/2003.


\bigskip


\begin{thebibliography}{31}


\bibitem{C1} M. Choptuik, Phys. Rev. Lett. {\bf 70}, 9 (1993)

\bibitem{G} C. Gundlach, Critical phenomena in gravitational collapse,  
gr-qc/0210101

\bibitem{R} M. Roberts, Gen. Relativ. Gravitation {\bf 21}, 907 (1989)

\bibitem{B} P. Brady, Class. Quant. Grav.{\bf 11}, 1255 (1995)

\bibitem{ONT} Y. Oshiro, K. Nakamura, A. Tomimatsu,
Prog. Theor. Phys. {\bf 91},1265 (1994)



\bibitem{SUS} R. Sussman, J. Math. Phys. {\bf 32}, 223 (1991)

\bibitem{BURKO} L. Burko, Gen. Relativ. Gravitation {\bf 29}, 259
(1997)

\bibitem{GAR} D. Garfinkle, Phys.Rev. D {\bf 63}, 044007 (2001)

\bibitem{CF} G. Clement, A. Fabbri, Nucl.Phys. B {\bf 630}, 269 (2002)


\bibitem{C2} S. Liebling, M. Choptuik, Phys. Rev. Lett. {\bf 77}, 1424 
(1996)

\bibitem{CS} T. Chiba, J. Soda, Prog. Theor. Phys. {\bf 96}, 567 (1996)

\bibitem{DOCT} H. de Oliveira, E. Cheb-Terrab, Class. Quant. Grav.{13}, 
425 (1996)

\bibitem{DO} H. de Oliveira, Self-Similar Collapse in Brans-Dicke Theory
and Critical Behavior, gr-qc/9605008

\bibitem{Y} S. Yazadjiev, Phys. Rev. D {\bf 65}, 084023 (2002)



\end{thebibliography}
\end{document}